\documentclass[preprintnumbers,amsmath,amssymb] {revtex4}
\usepackage{graphicx}
\usepackage[normalem]{ulem}
\usepackage[svgnames]{xcolor}
\usepackage{bm}
\usepackage{multirow}\usepackage{float}
\usepackage{titlesec}

\begin{document}
\title{Positive Seebeck Coefficient in Highly doped  La$_{2-x}$Sr$_x$CuO$_4$ ($x$=0.33); Its Origin and Implication}

\author{Hao Jin$^{1}$}
\thanks{Present address: CAS Key Laboratory of Nanosystem and Hierarchical Fabrication, CAS Center for Excellence in Nanoscience, National Center for Nanoscience and Technology, Beijing 100190, China}
\author{Alessandro Narduzzo$^{2}$}
\thanks{Present address: Physics Department, University of Bath, North Rd, Claverton Down, Bath BA2 7AY, UK }
\author{Minoru Nohara$^{3}$}
\author{Hidenori  Takagi$^{4,5}$}
\author{N. E. Hussey$^{2,6}$}
\author{Kamran Behnia$^{1}$}
\email{kamran.behnia@espci.fr}
\affiliation{(1) LPEM (ESPCI - CNRS - Sorbonne University), PSL University, 75005 Paris, France\\
(2) H. H. Wills Physics Laboratory, University of Bristol, Tyndall Avenue, Bristol, BS8 1TL, UK\\ 
(3) Research Institute for Interdisciplinary Science, Okayama University, Okayama 700-8530, Japan\\
(4) Max Planck Institute for Solid State Research, 70569 Stuttgart, Germany\\
(5)Department of Physics, University of Tokyo \\ Tokyo 113-0033, Japan\\
(6) High Field Magnet Laboratory (HFML-EMFL) and Institute for Molecules and Materials, Radboud University, 6525 ED Nijmegen, The Netherlands}

\date{\today}
\begin{abstract}
 We present a study of the thermoelectric (Seebeck and Nernst) response in heavily overdoped, non-superconducting La$_{1.67}$Sr$_{0.33}$CuO$_4$. In spite of the electron-like curvature of the Fermi surface, the Seebeck coefficient is positive at low temperatures. Such a feature, previously observed in copper, silver, gold and lithium, is caused by a non-trivial energy dependence of the scattering time. We argue that this feature implies a strong asymmetry between the lifetime of occupied and unoccupied states along the zone diagonals and such an electron-hole asymmetry impedes formation of Cooper pairs along the nodal direction in the superconducting ground state emerging at lower doping levels. 
\end{abstract}

\maketitle

Cuprates are a family of layered materials each with a Mott insulating parent which turns superconducting with a sizeable critical temperature upon doping~\cite{Lee2006}. Their normal state transport properties exhibit highly anomalous behaviour and a remarkable evolution with doping, temperature and magnetic field (for recent reviews, see \cite{Hussey2018,Proust2019}). In hole-doped cuprates, superconductivity fades gradually away beyond optimal doping $p \sim$ 0.16 before vanishing above a critical threshold $p_{SC} \sim 0.30$, though only in La$_{2-x}$Sr$_{x}$CuO$_4$ (LSCO) has this threshold ever been exceeded \cite{Nakamae2003,Cooper2009}.

\begin{figure}
\centering
\includegraphics[width=8.5 cm]{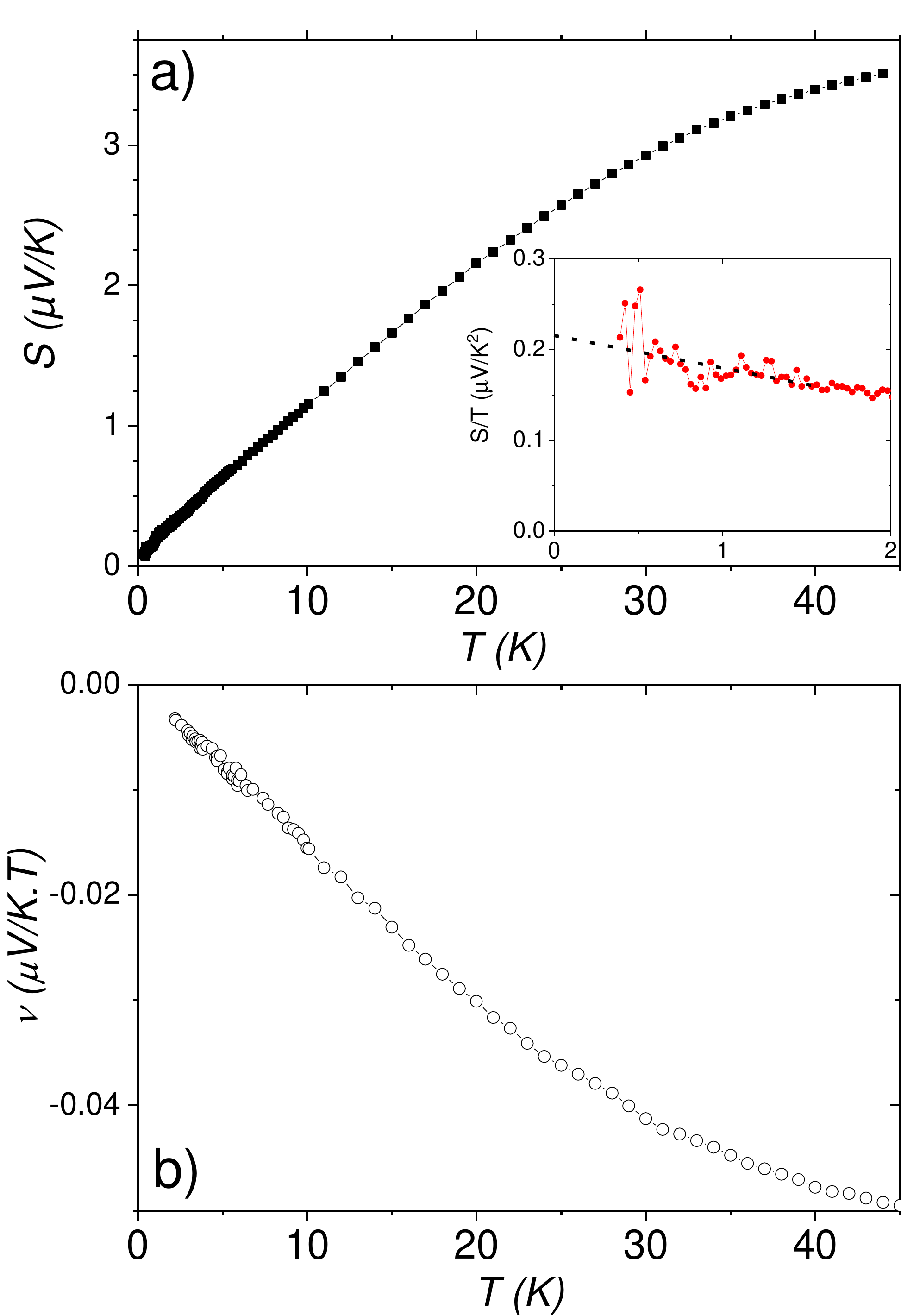}
\caption{\textbf{(Color online) Seebeck and Nernst coefficients:}  a) Temperature dependence of the in-plane Seebeck coefficient $S$, in La$_{1.67}$Sr$_{0.33}$CuO$_4$. Inset shows the low temperature $S/T$ showing the uncertainty in the zero-temperature slope. b) a) Temperature dependence of the in-plane Nernst coefficient $\nu$ measured at $\mu_0 H$ = 1 T.}
\label{fig:S-nu}
\end{figure}

In heavily overdoped La$_{1.67}$Sr$_{0.33}$CuO$_4$ (LSCO33), the in-plane resistivity $\rho_{ab}(T)$ follows the expected $T^2$ dependence of a correlated Fermi liquid below 50 K~\cite{Nakamae2003}. Removing carriers from this metal leads to the emergence of superconductivity as well as a 'strange' metal regime with a robust $T$-linear component in $\rho_{ab}(T)$ at low temperature~\cite{Cooper2009}. Scrutinizing the non-superconducting metal above $p_{SC}$ may provide clues for the origin of both.

In this paper, we present data on the thermoelectric response of LSCO33. In the zero temperature limit, both the Seebeck and Nernst coefficients are $T$-linear, as expected for a Fermi liquid. The sign and the slope of the Seebeck coefficient, however, point to a remarkably large lifetime asymmetry between occupied and unoccupied states. We argue that this impedes the formation of Cooper pairs along the zone diagonal ($\pi$, $\pi$) and thus provides the starting point for the emergence of a superconducting state with a $d_{x^2-y^2}$ symmetry.   

\begin{figure}
\centering
\includegraphics[width=8.5 cm]{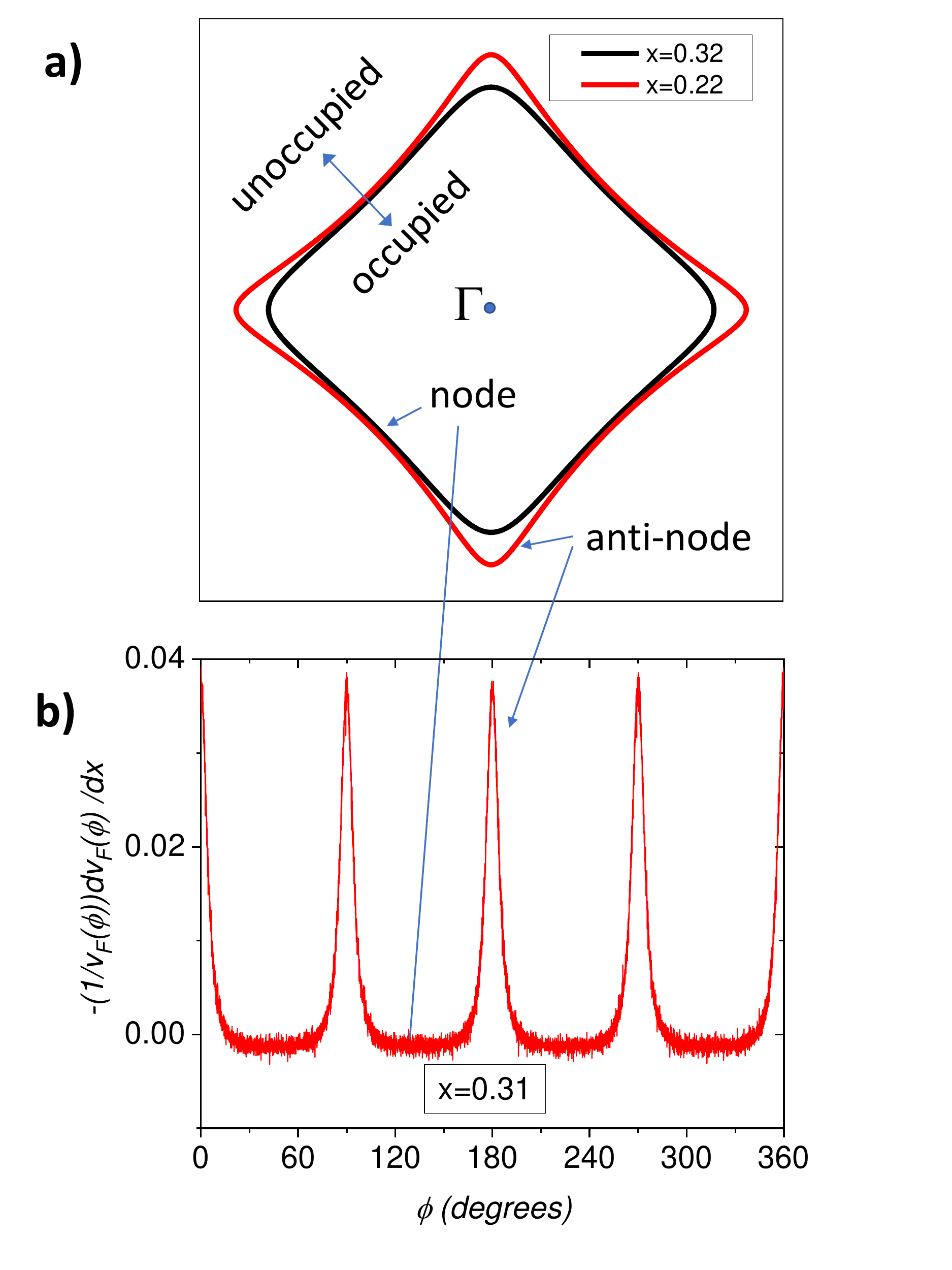}
\caption{\textbf{(Color online) The Fermi surface:} a) Tight binding Fermi surface of  La$_{2-x}$Sr$_{x}$CuO$_4$ for $x$ = 0.22 and $x$ = 0.32. The Fermi surface is electron-like. Note the large contrast between the doping evolution along nodal and anti-nodal directions.  b) The  relative change in Fermi velocity $v_F$ at azimuthal angle $\phi$ as one passes from $x$ = 0.30 to $x$ = 0.32. The change is arrested along the nodal direction and is maximal around the anti-nodes.}
\label{fig:FS}
\end{figure}

The resistivity of the single crystal measured in the present work showed no trace of superconductivity down to 95 mK  (see Ref.~\cite{Nakamae2003, Nakamae2009} for more details about the crystal growth, annealing conditions and characterization). The Seebeck and Nernst coefficients were subsequently measured (in 2004) using a standard one-heater-two-thermometers technique. Fig. \ref{fig:S-nu} shows their $T$-dependence between 0.4 K and 45 K. Both coefficients show a quasi $T$-linear dependence at low $T$. The asymptotic zero-temperature slope of the Seebeck coefficient is $S/T$ = +0.21 $\mu$V.K$^{-2}$, while for the Nernst coefficient, the slope is $\nu/T$ = -1.5 nV.T$^{-1}$.K$^{-2}$. 

The slope of the Nernst coefficient obtained in this measurement was first discussed in Ref.~\cite{Behnia2009}. There, it was argued that in the semi-classical picture, the amplitude of the slope is given by a set of fundamental constants ($\pi^2 k_B/3e$) multiplied by the ratio of the mobility $\mu_{\rm H}$ to the Fermi energy $E_F$. This picture is backed by available experimental data on numerous metals~\cite{Behnia2009,Behnia2016}. In the specific case of LSCO33, the low-$T$ $\nu/T$ is in fair agreement with estimates of $\mu_{\rm H} \approx 100$ (from magnetoresistance \cite{Nakamae2003}) and $E_F \approx$ 5900 K (from specific heat~\cite{Nakamae2003}). 

In this report, we will focus on the Seebeck response and discuss the significance and implications of both its sign and amplitude. The Seebeck coefficient in cuprates has been the subject of numerous studies  (See chapter 8 in Ref.~\cite{Behnia2015b} for a review). In the case of LSCO, it was previously studied in single crystals (up to $x$ = 0.3 by Nakamura and Uchida~\cite{Nakamura1993}) and in polycrystalline powders (up to $x$ = 0.45 by Cooper and Loram ~\cite{Cooper1996} and up to $x$ = 0.35 by Elizarova and Gasumyants ~\cite{Elizarova2000}). Our LSCO33 data, which extends down to sub-kelvin temperature, agrees well with these earlier studies in the overlapping temperature ranges. Our data also smoothly connects to what was recently reported by Collignon \textit{et al.} in Eu-substituted  LSCO for $x<0.26$~\cite{collignon2020thermopower}.

The first remarkable fact about the Seebeck coefficient of LSCO33 is its positive sign. Angle resolved photoemission spectroscopy (ARPES)~\cite{Yoshida_2007,Yoshida2006,Horio2018} studies have extensively documented the emergence of an electron-like Fermi surface in LSCO for $x > 0.2$, i.e. closed around the $\Gamma$ point in the Brillouin zone. Fig.~\ref{fig:FS} shows the Fermi surface derived from a tight binding model with nearest-neighbor hopping parameters chosen to fit the ARPES-resolved Fermi surface~\cite{Yoshida2006,Horio2018}. Thus, given the electron-like character of the Fermi surface, one would naively expect the thermopower of LSCO33 to be negative.

Interestingly, the Hall coefficient of LSCO33 is also positive~\cite{Narduzzo2008}. This observation was explained by taking into account both the curvature of the Fermi surface and the strong angle dependence of the scattering time $\tau$ and mean-free-path of the mobile carriers at the Fermi level~\cite{Narduzzo2008}. For the Seebeck coefficient, as we will see below, it is the \textit{energy} dependence of $\tau$ which matters. 

In numerous Fermi liquids, there is an empirical correlation between the slope of the diffusive Seebeck coefficient $S/T$ and the magnitude of the electronic specific heat $\gamma$~\cite{Behnia_2004}. A dimensionless ratio of these two quantities can be defined using Avogadro's number $N_{Av}$ and the charge of an electron, $e$:
\begin{equation}
    q=\frac{SN_{Av}e}{T\gamma}
\end{equation}

In dense Fermi liquids (i.e. those with roughly one mobile electron per formula unit), $q$ is of order unity. This observation, first reported in 2004~\cite{Behnia_2004} has been confirmed in numerous cases. The strength of correlation among the conduction electrons tunes $\gamma$ over several orders of magnitude ($\approx$ 1-1000 mJ.K$^{-2}$.mol$^{-1}$). Concomitantly, it modifies the absolute value of $S/T$. Since both these quantities track the entropy accumulated by electrons, such a correlation may not be surprising, though its persistence in many multi-band metals with a Fermi surface consisting of pockets of both signs remains a puzzle.

In LSCO33, where $S/T$ = +0.21 $\mu$V.K$^{-2}$ and $\gamma$ = 6.9 mJ.K$^{-2}$.mol$^{-1}$~\cite{Nakamae2003}, one finds $q=+2.8$. In dilute Fermi liquids, $q$ can be significantly larger than unity, because the entropy per volume lags behind entropy per carrier. In URu$_2$Si$_2$, for example, $|q| \simeq 11$~\cite{Zhu2009}. LSCO33, on the other hand, is a dense Fermi-liquid with 1.3 carriers per formula unit. Hence, not only the sign, but also the enhanced value of $q$ demand an explanation. 

As it turns out, this is not the first example of a system with a large simple Fermi surface -- occupying more than half the Brillouin zone -- exhibiting an anomalous Seebeck coefficient. In noble metals (Cu, Ag, Au), the Seebeck coefficient is positive and $T$-linear well above the phonon drag peak~\cite{MacDonald}, despite their Fermi surfaces being electron-like ~\cite{shoenberg2009}. In 1967, Robinson called this puzzle \lq a nagging embarrassment to the theory of the ordinary electronic transport properties of solids'~\cite{Robinson1967}.  Interestingly, the $q$ ratio is +0.75 in Cu, +0.81 in Ag and +0.86 in Au~\cite{Behnia2015b}, i.e. $S/T$ has the right magnitude, just the wrong sign.

This \lq reversed sign thermopower' puzzle in Cu, Ag, Au and Li (the alkali metal also showing an unexpected positive Seebeck response) has been addressed by Robinson~\cite{Robinson1967} and more recently by Xu, Di Gennaro and Verstraete~\cite{Xu2014,Xu2020}. Robinson argued that a mean-free-path rapidly decreasing with increasing energy would provide a solution to the puzzle and this can arise due to the structure of the electron-ion pseudopotential. Xu ~\textit{et al.} carried out first-principle calculations and found that in Li, the sign reversal is driven by density and lifetime asymmetries between states above and below the Fermi level, $E_F$~\cite{Xu2014}. Similar calculations for noble metals also produced a positive sign due to a non-trivial asymmetry in electron-phonon coupling for electronic states at the two sides of the chemical potential~\cite{Xu2020}. These results motivate us to search for a similar solution for the puzzle of  the \lq wrong' sign in LSCO33. 

\begin{figure}
\centering
\includegraphics[width=8.5 cm]{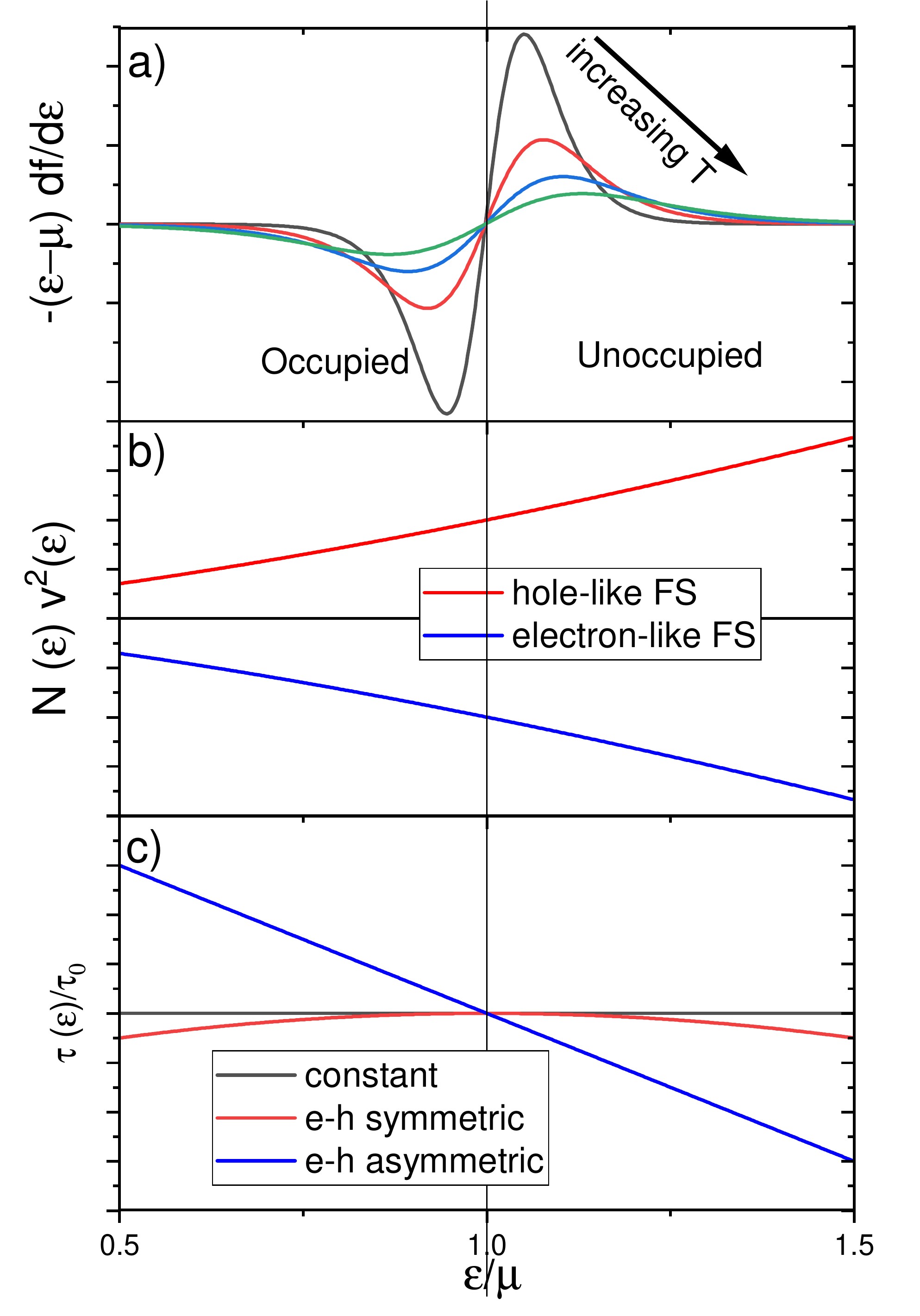}
\caption{\textbf{(Color online) Inverting the sign of the Seebeck response: } The Seebeck coefficient is the result of the integration of three components over the Fermi surface. These three ingredients, sketched as a function of energy normalized to the chemical potential $\mu$, are: a) The pondering function; b) The product of  density of states and the square of velocity of a gas of free electrons (blue) and free holes (red): c) The scattering time $\tau(\epsilon)$ in three distinct scenarios. When $\tau(\epsilon)$ is constant (black) or e-h symmetric (red), the sign of the Seebeck coefficient remains positive for holes and negative for electrons. When the energy dependence is such that the unoccupied states are significantly more scattered than the occupied states (blue) the sign will be inverted.}
\label{fig:FD}
\end{figure}

The Seebeck coefficient is defined as the ratio of thermoelectric conductivity $\alpha$ to the electric conductivity $\sigma$. The Boltzmann equation links both coefficients to $\tau$, the density of states $N(\epsilon)$ and the velocity $v$ ~\cite{ziman_1972,Behnia2015b}:

\begin{equation}
 \alpha=  -e \int \tau(\epsilon_k)v(\epsilon_k).v(\epsilon_k)N(\epsilon_k)\frac{(\epsilon_k-\mu)}{T}\frac{\partial f}{\partial \epsilon_k}  \,d\epsilon_k 
\end{equation}

\begin{equation}
 \sigma=  -e^2 \int \tau(\epsilon_k)v(\epsilon_k).v(\epsilon_k)N(\epsilon_k)\frac{\partial f}{\partial \epsilon} \,d\epsilon_k 
\end{equation}

Here, the integrals are over the whole Fermi surface, $f$ is the Fermi-Dirac distribution and $\mu$ is the chemical potential. The expression for $\alpha$, contains a material-independent pondering factor together with material-dependent parameters. As seen in  Fig.~\ref{fig:FD}a, the pondering factor is anti-symmetric about the chemical potential. In the absence of electron-hole asymmetry near the chemical potential, $\alpha$ would be zero. However, even in a free electron gas, such asymmetry exists; both the velocity ($v(\epsilon_k) \propto \sqrt{\epsilon_k}$) and the density of states ($N(\epsilon_k) \propto \sqrt{\epsilon_k}$) grow with energy (See Fig.~\ref{fig:FD}b). As a result, $\alpha$ of a free electron (hole) gas is negative (positive). Such a correspondence between the sign of the Seebeck coefficient and the sign of carriers survives even in more complex Fermi surface geometries provided that the energy dependence of the scattering time (or the mean-free-path) does not alter the result. Note that Fermi's golden rule, by linking the scattering rate and the density of states, implies that features in $N(\epsilon_k)$ will have counterparts in $\tau(\epsilon_k)$. Note that $v$, $\tau$ and $N(\epsilon_k)$ can all have significant momentum dependence too. As we shall see below, this $k$-space anisotropy also plays a prominent role here.

Fig. \ref{fig:FD}c shows three possible scenarios for the energy dependence of the scattering time. In the first two cases, $\tau$ is constant or its energy dependence is symmetric as one moves off the chemical potential and there is no effect on the sign of $S$. In the third case, however, $\tau$ decreases sufficiently rapidly with increasing energy that it inverts the overall balance of the responses of the occupied and unoccupied states, as originally shown by Robinson~\cite{Robinson1967}. 
 
Such an energy dependence can arise for different reasons. According to first principle calculations on Li~\cite{Xu2014,Xu2020}, a feature in density of states just below the chemical potential skews the available phase space for scattering~\cite{Xu2014}. In copper, on the other hand, the density of states is flat near the chemical potential~\cite{Xu2020}, and it is the electron-phonon coupling that is energy dependent. Robinson's phenomenological model -- invoking a screened electron-ion pseudopotential -- leads to a similar conclusion~\cite{Robinson1967}. 

Coming back to LSCO33, an energy-dependent scattering time would provide a natural explanation for the positive sign of the Seebeck coefficient, though the large magnitude of $q$ implies that the hole-particle asymmetry may be even more pronounced than in noble metals. If one assumes a conventional energy dispersion for a free electron gas but with an energy dependence of the mean-free-path that follows a power law: $\ell \propto \epsilon ^ {-\delta} $,  then $\delta$ would be set by $3q/2+1$. In copper, for example, an inverted $q=+0.75$ would then require  $\delta \approx 2.1$~\cite{Behnia2015b}, while an inverted $q=+2.8$ would require $\delta$ exceeding 5. This may suggest that in our present case, the energy dependence of the mean-free-path is stronger than a simple power law. 

The evolution of the Fermi surface with doping shown in Fig. \ref{fig:FS} indicates a very plausible nexus for a strong particle-hole asymmetry. The introduction of additional holes between $x$ = 0.22 and $x$ =0.32, shifts the Fermi surface almost exclusively along the anti-nodal direction and not at all along the zone diagonal. This implies that along the nodal orientation the density of states does not smoothly grow as a function of the chemical potential and therefore, the phase space for scattering to unoccupied states above the chemical potential is extremely reduced. As seen in Fig. \ref{fig:FS}b, the doping-induced change of the Fermi velocity along the zone diagonal is negligible in comparison with the anti-nodal direction. This indicates that the Seebeck coefficient is dominated by the contribution of nodal quasi-particles with strong asymmetry in the lifetime between occupied and unoccupied states. 

The dichotomy between the nodal and anti-nodal contributions to transport has been demonstrated by angle-dependent magnetoresistance (ADMR) studies, first in Tl-2201~\cite{HUssey2003,Abdel-Jawad2006} and more recently in Eu-LSCO~\cite{Grissonnanche2020}. However, the focus of attention of both studies was the $T$-dependence of the anisotropic $\tau$ and not the energy dependence of $\tau$ and its anisotropy.  What causes the scattering time to be strongly energy dependent along the nodes remains a question to be addressed by microscopic theory. At this stage, let us point out that this identification has a possible link with the superconducting gap structure and pairing symmetry.

LSCO33 becomes a superconductor upon removal of mobile carriers. At the same time, a $T$-linear scattering rate emerges below $p_{SC}=0.3$. The finite positive $S/T$, on the other hand, does not appear to be affected by the emergence of superconductivity. Indeed, the magnitude of $S/T$ in non-superconducting LSCO33 is almost identical with what has been recently found by Collignon \textit{et al.} in superconducting Eu-LSCO for $0.23<x<0.26$ ($\approx$ +0.2 $\mu$V. K$^{-2}$)~\cite{collignon2020thermopower}.

In a BCS superconductor, Cooper pairs are formed by the superposition of states above and below the Fermi level within an energy window of the size of the gap~\cite{Tinkham1996}. Strong asymmetry between quasiparticle lifetimes across the occupation boundary would impede the formation of Cooper pairs. Our analysis finds that in the nodal orientation, the density of states and the quasi-particle lifetime do not evolve smoothly across the Fermi energy. Such an electron-hole asymmetry along ($\pi$, $\pi$) would obliterate the superconducting gap along the nodes, even if the attractive interaction leading to the formation of Cooper pairs were isotropic. It remains to be seen how the $d_{x^2-y^2}$ pairing symmetry of the superconductor~\cite{Tseui2000} and the anisotropic energy-dependent scattering phase space of the strange metal connect to each other.

KB is supported by the Agence Nationale de la Recherche  (ANR-18-CE92-0020-01; ANR-19-CE30-0014-04). NEH is supported by the Netherlands Organisation for Scientific Research (NWO) (Grant
No. 16METL01)—“Strange Metals” and by the European Research Council (ERC) under the European Union's Horizon 2020 research and innovation programme (Grant Agreement No. 835279-Catch-22).

\bibliography{biblio}
\end{document}